\documentclass{emulateapj}
\shorttitle{Temperature of Starless Cores}
\shortauthors{Schnee, S. et al.}

\begin{document}

\newcommand{\myemail}{schnee@astro.caltech.edu}
\newcommand{\amm}{NH$_3$}
\newcommand{\kms}{km s$^{-1}$}
\newcommand{\nh}{N$_{\rm H}$}
\newcommand{\td}{T$_{\rm d}$}
\newcommand{\tc}{T$_{\rm kin}$}
\newcommand{\tco}{T$_{\rm CO}$}
\newcommand{\sigamm}{$\sigma_{\rm NT}$}
\newcommand{\sigco}{$\sigma_{\rm CO}$}
\newcommand{\AV}{$A_V$}
\newcommand{\ccm}{cm$^{-3}$}

\title{The Gas Temperature of Starless Cores in Perseus} 
\author{S. Schnee$^1$, E. Rosolowsky$^{2,3}$, J. Foster$^2$, M. Enoch$^4$, A. 
Sargent$^1$}
\affil{$^1$Division of Physics, Mathematics and Astronomy, 
       California Institute of Technology, 770 South Wilson Avenue,
       Pasadena, CA 91125 \\
       $^2$Harvard-Smithsonian Center for Astrophysics, 60 Garden
       Street, Cambridge, MA 02138 \\
       $^3$University of British Columbia Okanagan, 3333 University
       Way, Kelowna, BC V1V 1V7, Canada \\
       $^4$Department of Astronomy, University of California, Berkeley, 
       CA 94720}
\email{\myemail}

\begin{abstract}

In this paper we study the determinants of starless core temperatures
in the Perseus molecular cloud.  We use \amm\ (1,1) and (2,2)
observations to derive core temperatures (\tc) and data from the
COMPLETE Survey of Star Forming Regions and the c2d Spitzer Legacy
Survey for observations of the other core and molecular cloud
properties.  The kinetic temperature distribution probed by \amm\ is
in the fairly narrow range of $\sim$ 9 - 15 K.  We find that cores
within the clusters IC348 and NGC1333 are significantly warmer than
``field'' starless cores, and \tc\ is higher within regions of larger
extinction-derived column density.  Starless cores in the field are
warmer when they are closer to class O/I protostars, but this effect
is not seen for those cores in clusters.  For field starless cores,
\tc\ is higher in regions in which the $^{13}$CO linewidth and the
1.1mm flux from the core are larger, and \tc\ is lower when the the
peak column density within the core and average volume density of the
core are larger.  There is no correlation between \tc\ and $^{13}$CO
linewidth, 1.1mm flux, density or peak column density for those cores
in clusters.  The temperature of the cloud material along the line of
sight to the core, as measured by CO or far-infrared emission from
dust, is positively correlated with core temperature when considering
the collection of cores in the field and in clusters, but this effect
is not apparent when the two subsamples of cores are considered
separately.

\end{abstract}


\section{Introduction}

Starless cores are the link between molecular clouds and Class 0
protostars. They are smaller and denser (r $\sim$ 0.1 pc, n $\gtrsim
10^4$ cm$^{-3}$) than, and kinematically distinct from, the molecular
clouds in which they are found \citep{Goodman98}.  Because starless
cores have no source of internal heating, the temperature of a
starless core is determined by external illumination and
self-shielding.  The temperatures of starless cores are set by the
interstellar radiation field, and in some cases luminous nearby stars,
attenuated by the molecular cloud.  The temperature profile within a
starless core is determined by the heating of its outer layers and the
internal extinction, making the centers of starless cores colder than
their outer layers \citep{Evans01, Stamatellos07, Zucconi01}.

The temperatures of starless cores can be inferred from thermal
emission from dust or from molecular lines.  Deriving core
temperatures from continuum emission is complicated by uncertainties
in the dust emission spectrum.  For instance, in cold and dense
regions dust grains are expected to coagulate and become covered in
icy mantles, creating variation in the emissivity spectral index
\citep{Ossenkopf94}.  In addition, below a density of $\sim3 \times
10^4$ the dust temperature is not coupled to the gas temperature
\citep{Goldsmith01, Galli02}, which is the more important quantity
because the gas makes up $\sim$99\% of the mass.  One of the best
tracers of gas temperature is the rotation-inversion lines of \amm.
Ammonia is particularly well suited for this type of study because 1)
\amm\ does not seem to deplete onto dust grains in starless cores, and
in fact may have an enhanced abundance where CO is depleted
\citep{Tafalla02} 2) \amm\ has hyperfine splitting of its inversion
lines, making fits to the temperature, velocity and optical depth more
certain 3) the \amm\ (1,1) and (2,2) transitions are separated by a
small difference in frequency, so it is possible to simultaneously
observe both sets of lines.  In this paper we use ammonia (1,1) and
(2,2) observations \citep{Rosolowsky08} to derive the temperature of
starless cores.  We examine whether these data support the several
models of the relationship between core properties and the molecular
cloud where they are found \citep[e.g.,][]{Evans01,
Bethell04,Stamatellos04}

Perseus is one of the Gould's belt molecular clouds included in the
Spitzer legacy project ``From Molecular Cores to Planet-Forming
Disks'' (c2d) \citep{Evans03} and the COMPLETE survey of star-forming
regions \citep{Ridge06}, at a distance of 250 pc \citep{Cernis03,
Belikov02}.  Because it is so well observed, there is a wealth of data
on the cores and the cloud that enables us to study the influences of
various internal and environmental factors on the temperature of 50
starless cores in Perseus.

\section{Data}

\subsection{Far-Infrared} \label{OBSFIR}

As described in \citet{Schnee08}, we used far-infrared emission maps
at 70, 100 and 160 \micron\ to derive the dust temperature and column
density of the Perseus molecular cloud.  The 70 and 160 \micron\ maps
were obtained as part of the c2d survey \citep{Rebull07}, and the 100
\micron\ map was taken from the IRIS reprocessing of the IRAS all-sky
maps \citep{Miville05}.  The resolution of the emission-derived column
density (\nh) and dust temperature (\td) maps with which we will
compare the \amm\ temperature (\tc) is 40\arcsec, as set by the 160
\micron\ data.

The dust temperature is calculated from the ratio of the observed flux
at 70 and 160 \micron, taking into account the color correction at
both wavelengths and the emission from transiently heated very small
dust grains at 70 \micron.  The column density is derived from the
flux at 160 \micron\ and the dust temperature, allowing for a variable
dust emissivity and normalized to the extinction measured from NIR
reddening.  Maps of the emission-derived dust temperature and column
density, and the procedure used to make the maps, are presented in
detail in \citet{Schnee08}.

\subsection{Near-Infrared} \label{OBSNIR}

Data from the final release of the Two Micron All Sky Survey (2MASS)
point source catalog was used to create a large-scale extinction map
of Perseus as part of the COMPLETE survey \citep{Ridge06}.  The NICER
algorithm estimates the column density towards background stars by
comparing their observed colors with the (assumed) instrinsic colors
\citep{Lombardi01}.  Although we expect the extinction map to be our
least-biased estimate of the cloud column density \citep{Schnee06,
Goodman08}, its 5\arcmin\ resolution is roughly ten times coarser than
our other maps, which smooths out density gradients and makes it
somewhat harder to compare the cloud and core properties.

\subsection{\amm\ (1,1) and (2,2)} \label{OBSNH3}

We used observations of 193 dense cores and core candidates in the
Perseus Molecular Cloud using the 100-m Robert F. Byrd Green Bank
Telescope (GBT) as described in \citet{Rosolowsky08}. For each target,
single-pointing (31\arcsec\ FWHM) observations were made of the \amm\
(1,1) and (2,2) lines with 0.024 \kms\ spectral resolution.  The
targets for the ammonia survey were drawn, in order of precedence,
from 1) the locations of millimeter cores in the Bolocam survey of the
region \citep{Enoch06} 2) the locations of submillimeter cores in the
SCUBA survey of the region \citep{Kirk06}, 3) sources in the
literature survey of \citet{Jijina99}, and 4) cold,
high-column-density objects in the dust map produced by
\citet{Schnee08}, and 5) sources identified by eye that appear in both
the Bolocam and SCUBA maps but were not included in the published
catalogs.  The starless cores discussed in this paper were drawn from
from only the first two categories.

The kinetic temperatures of the starless cores (Section \ref{IDENT}
describes how we identify which cores are starless) are determined by
a simultaneous fit to the (1,1) and (2,2) emission profiles,
accounting for collisional deexcitation of the two states. The same
fit also yields the width of the \amm\ line, which is decomposed into
its thermal and non-thermal components based on the temperature
determination.  Although there are likely gas temperature and density
gradients within each starless core on both theoretical and
observational grouds \citep{Crapsi07}, the fitting of the
ammonia-derived temperature is essentially an average over the core
properties.  However, a single-temperature model with uniform
excitation conditions is a reasonably good fit to nearly all of the
starless core in this sample.  The NH$_3$ observations can be used to
accurately derive the kinetic temperatures of starless cores with
$8\mbox{ K}\le T_{\mathrm{kin}} \le 40\mbox{ K}$ where the lower and
upper limits are set by sensitivity and transition thermalization
respectively \citep{Walmsley83}.  Further details about the
observations and data reduction are described in \citet{Rosolowsky08}.

\subsection{$^{12,13}$CO (1--0)} \label{OBSCO}

Observations of the $^{12}$CO (1--0) and $^{13}$CO (1--0) transitions
at 115 and 110 GHz were carried at the 14-m Five College Radio
Astronomy Observatory (FCRAO) telescope as part of the COMPLETE
survey.  The data have 0.07 \kms\ spectral resolution and 40\arcsec\
spatial resolution.  The data reduction and mapping techniques are
described in \citet{Ridge06}. The CO maps cover approximately the same
area in Perseus as the c2d survey, giving us a good estimate of the
kinematics of the ambient, less dense, material in the molecular
cloud.  \citet{Pineda08} use this data set to determine the gas
temperature and column density in regions in which the $^{13}$CO
(1--0) transition is optically thin (the $^{12}$CO line is assumed to
be optically thick everywhere) and the gas is not subthermally
excited.

\subsection{1.1mm Continuum Maps} \label{OBSBOLO}

As part of a survey that is complemetary to c2d, \citet{Enoch06}
mapped 7.5 deg$^2$ of the Perseus molecular cloud at 1.1mm with
Bolocam at the Caltech Submillimeter Observatory.  The resolution is
31\arcsec, which matches very well with our \amm\ observations, and
all 122 1.1mm-detected cores were observed with the GBT.  The Bolocam
observations are most sensitive to dense cores ($n \gtrsim 2 \times
10^4$ cm$^{-3}$) with density contrasts at least 30 times greater than
the background cloud, which matches the description of a starless core
embedded in a molecular cloud quite well \citep{Bergin07}.  In addition
to using the 1.1mm data for target selection, we use the 1.1mm derived
fluxes and sizes to compare the \tc\ with core properties such as
mass, column density and volume density.

\section{Identification of Starless Cores} \label{IDENT}

To ensure that we only study the starless core population in Perseus,
we compare the positions of our \amm\ observations with the
coordinates of known protostars, as found through the analysis of
\citet{Enoch08}, \citet{Jorgensen07} and \citet{Hatchell07}.
\citet{Jorgensen07} identify 49 deeply embedded Young Stellar Objects
(YSOs) in Perseus using Spitzer/IRAC color cuts, Spitzer/MIPS and 850
\micron\ SCUBA correlations, and SCUBA morphologies.
\citet{Hatchell07} uses NIR/MIR (2MASS/IRAC) detections and/or
detections of protostellar outflows to separate starless cores from
protostars, and find 56 protostars.  \citet{Enoch08} identifies 55
protostellar cores out of the total 122 millimeter cores in Perseus
using the spatial coincidence between 1.1mm continuum cores and
sources identified in the c2d Delivery Document \citep{Evans07,
Harvey07} as ``YSOc'' (YSO candidate), ``red'' or that are 70 \micron\
point sources and are not classified as ``Galc'' (Galaxy candidate).
In this paper we only consider those 50 ammonia pointings that are
called ``starless'' by the various definitions of all three papers, or
that are more than 30\arcsec\ from any identified protostar in any of
the three papers.  None of the cores we classify as ``starless'' are
identified as a low-luminosity embedded protostar by \citet{Dunham08}.
We also require that the starless cores have solid \amm-derived
temperatures and linewidths, meaning that \tc\ is greater than 2.73 K,
the \amm\ (2,2) detection is $> 5\sigma$ and the model and observed
spectra fit well by eye.  Of the 193 observations in the \amm\ survey,
137 of the pointings are not coincident with an identified protostar,
and 50 of these satisfy the signal to noise and goodness of fit
requirements.

\section{Analysis} \label{ANALYSIS}

Here we compare the ammonia-derived temperature with several
properties of the starless cores and their environment.  We considered
only those 50 starless cores with well-defined \tc\ and linewidth, so
we are able to derive their non-thermal linewidths.  All were detected
in the 1.1mm survey of \citet{Enoch06}, so for each we have the total
flux, peak flux (from which we derive the peak column density), mass
(from the total flux and the \amm\ temperature) and mean particle
density.  The FIR-derived dust temperature and column density are
considered to be believable as long as they fall within the range $12
\le T_d \le 20$ and $0 \le A_V \le 24.5$, which is true except in
those regions where the 160 \micron\ map is saturated or corrupted by
a nearby bright source.  There are 46 starless cores with believable
FIR-derived \td\ and \AV, and all 50 starless cores fall within the
area of our NIR-derived extinction map. The gaussian fitting algorithm
failed to converge when calculating the $^{13}$CO linewidth of \tco\
for 6 cores.  The internal properties of our sample of starless cores
are listed in Table \ref{COREPROPTAB}, and the properties of Perseus
cloud along the line of sight to the cores are given in Table
\ref{COREENVIRTAB}.  Although the distance from each starless core to
the nearest protostar found in the surveys of \citet{Enoch08,
Jorgensen07} and \citet{Hatchell07} is known, in our discussion of
possible correlations we consider only those with separations of 0.25
pc or smaller.

\subsection{\tc} \label{TC}

\begin{figure} 
\epsscale{1.0} 
\plotone{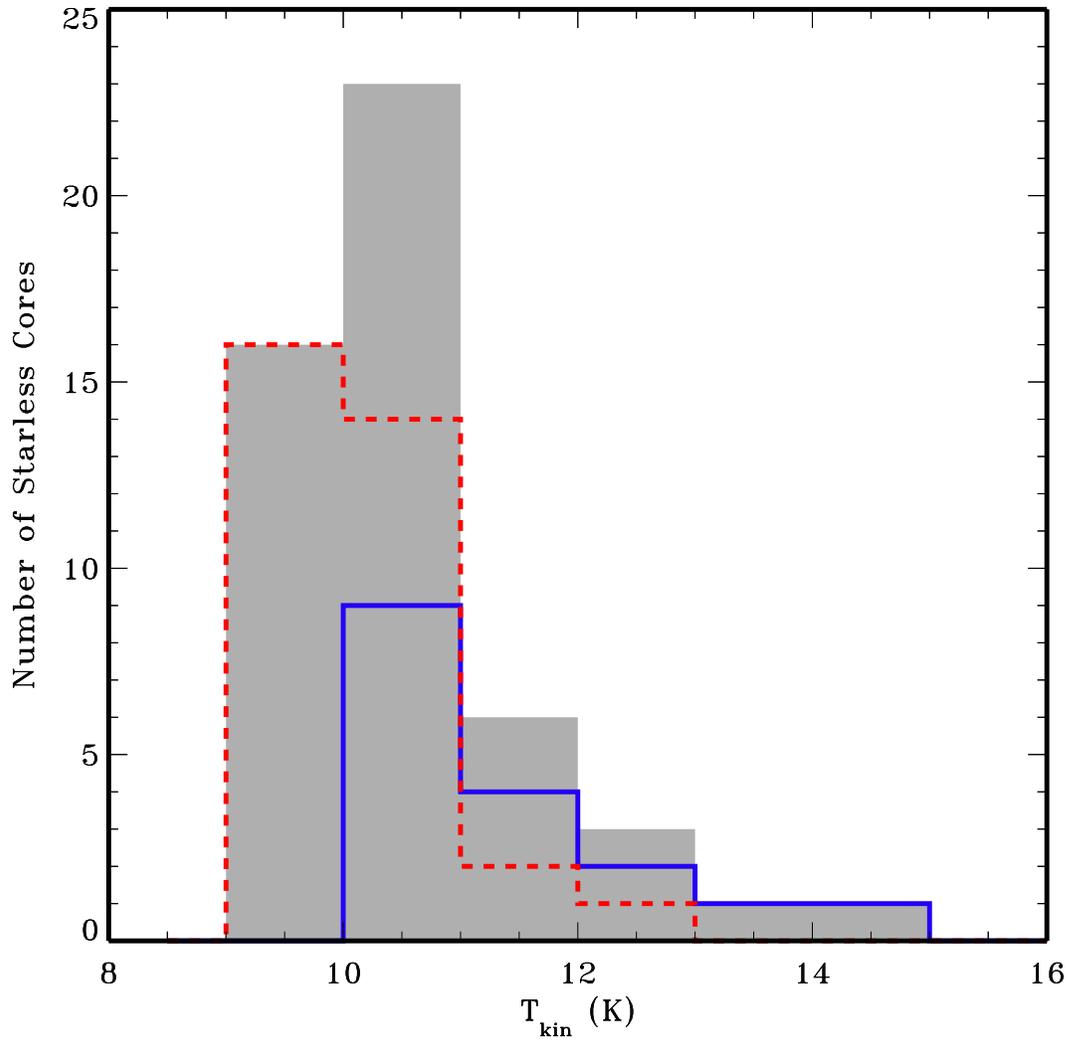} 
\caption{Histogram of \tc.  The filled (grey) histrogram is the
  temperature distribution of all 50 of the starless cores in Perseus
  for which we have well-fit \amm\ spectra.  The dashed (red) and
  solid (blue) histograms are the temperature distribution of the
  starless cores in the field and inside the clusters, respectively.
\label{TKINHIST}}
\end{figure}

The kinetic temperature distribution probed by \amm\ is quite narrow,
from $\sim$ 9 - 15 K, as seen in Fig.~\ref{TKINHIST}.  This may be due
to the relatively low angular resolution of the GBT pointings, whose
31\arcsec\ beam corresponds to probing the inner 0.019 pc (3.9$\times
10^3$ AU) of each core.  Models of externally heated prestellar cores
show that they are only significantly colder than 9 K within the inner
$\sim 10^3$ AU or warmer than 14 K outside a radius of $\sim 10^4$ AU,
with variation depending on the strength of the ISRF incident upon the
core \citep{Evans01, Stamatellos04}.  We are unlikely to be biased by
chemical effects, given that \amm\ is not seen to deplete in starless
cores \citep[e.g.,][]{Tafalla02}.  The critical density of \amm, $\sim
10^4$ \ccm\ \citep{Swade89}, is low compared the typical gas densities
observed in the cores in our sample ($10^4$ \ccm\ $\le n \le 10^6$
\ccm), so we expect the \amm\ observations to sample the full volume
of our cores.

The range of temperatures seen in the Perseus starless cores is
typical of those seen in other molecular clouds.  For instance,
\citet{Kirk07} derived dust temperatures of pre-stellar cores in
Taurus and Ophiuchus and found that most cores fall within the 9-13K
range.  The peak of the \tc\ distribution shown in Figure
\ref{TKINHIST} is between 10-11K, which agrees well with the median
temperature (11 K) derived from \amm\ observations of the 36 starless
as observed and identified by \citet{Jijina99} across many nearby star
forming regions.  The lower and upper quartiles of the temperatures of
the \citet{Jijina99} starless core sample (9.7 K and 14.0 K) are also
a close match to that shown in Figure \ref{TKINHIST}.

Numerical simulations of starless cores have been used to quantify how
their temperature profile is affected by variations in many
parameters, such as the strength of the ISRF, the density profile, and
the amount of extinction provided by the molecular cloud in which the
cores are embedded \citep[e.g.][]{Evans01, Galli02, Stamatellos03,
Stamatellos04}.  \citet{Evans01} calculate that for Bonnor-Ebert
spheres with central densities varying by about a factor of 100 (from
10$^5$ to 10$^7$ cm$^{-3}$), the central temperature varies by a
factor of $\sim$2 (6.5 - 10 K), but the temperature at a radius of
10$^{3.5}$ AU (the radius of the GBT beam at the distance of Perseus)
only varies by $\sim$1 K (9 - 10 K).  A doubling of the strength of
the ISRF results in a 1 K increase of the core temperature file over
the range of radii probed by the \amm\ observations in this study
\citep{Evans01}, while a factor of 100 increase in the ISRF would
result in an increase in the core temperature by a factor of $\sim$2
\citep{Galli02}.  Variations in the dust opacity also affect the
temperature of starless core, but \citet{Goncalves04} show that this
effect is quite small ($< 1$ K) for the range of dust properties that
one expects in these cores.  Given that the temperature of a
prestellar core (at a radius of 10$^{3.5}$ AU) is not especially
sensitive to its central density, the dust opacity, the strength of
the ISRF or the degree of extinction for the range of values that we
expect in Perseus, the narrow width of the temperature distribution
shown in Figure \ref{TKINHIST} and its peak at 10 K are roughly what
one would expect.

\subsection{\tc\ vs.~Location} \label{TGDIST}

The presence of local star formation appears to affect \tc.
Fig.~\ref{TKINHIST} shows the temperature distribution of all the
starless cores in Perseus (filled grey histogram), as well as that for
starless cores in the clusters NGC1333 and IC348 (open solid
histogram) and field starless cores (open dashed histogram).  The
boundaries for NGC1333 and IC348 are taken from \citet{Jorgensen07}.
It is clear that those cores in clusters are warmer than those outside
clusters.  The probability that the temperature distributions of
starless cores inside and outside clusters have the same mean is
$\simeq 10^{-5}$, as calculated by a Wilcoxon Rank-Sum Test, Student's
T-statistic or two-sided Kolmogorov-Smirnov test.

There are reasons to expect that starless cores in clusters would be
warmer than those outside clusters.  For instance, a cluster of
protostars may directly heat their environment.  In addition,
protostars inject turbulence into the surrounding medium, making it
more porous and less effective as a shield from the ISRF.  In
addition, \citet{Jijina99} find that dense cores that are not
associated with {\it IRAS} sources are warmer when found inside
clusters than outside.

\begin{figure}
\epsscale{1.0}
\plotone{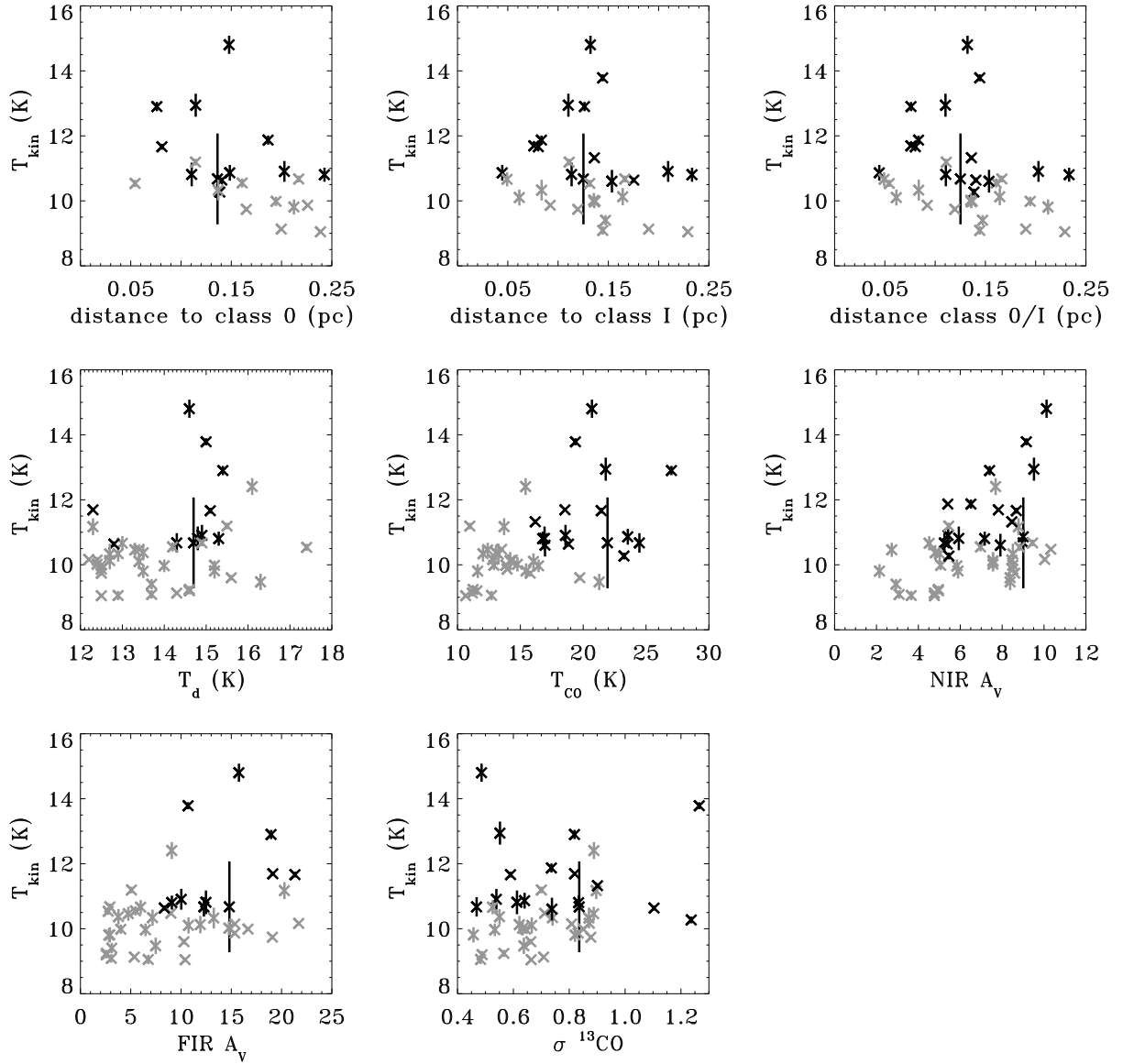}
\caption{\tc\ plotted against properties of the enviroment around the
  starless cores: (top left) distance to nearest Class 0 protostar,
  (top middle) distance to nearest Class I protostar, (top right)
  distance to nearest Class 0 or Class I protostar (middle left) dust
  temperature, (center) gas temperature, (middle right) column density
  of the molecular cloud derived from NIR extinction, (bottom left)
  column density of the cloud derived from FIR emission and (bottom
  middle) $^{13}$CO line width (1-D velocity dispersion).  In this
  plot Class 0 cores and Class I cores are identified as in
  \citet{Enoch08b}.  Starless cores in NGC1333 and IC348 are plotted
  in black, and field starless cores are plotted in grey. The
  1$\sigma$ errors in the fit of \tc\ are plotted for each
  core.\label{TKINENVIRONMENT}}
\end{figure}

\citet{Stamatellos07} predict that the presence of a nearby Class 0
protostar should not affect the temperature of a nearby starless core
at all, and that a nearby Class I protostar will only heat the outer
layers of a core, and only by $\sim1-2$K.  We test this by comparing
the \amm-derived temperatures of starless cores in Perseus to the
distance to the nearest Class 0 and I protostars (as determined by
\citet{Enoch08}).  As shown in Fig.~\ref{TKINENVIRONMENT} and Table
\ref{CORRELTAB}, we confirm that there is no significant correlation
between \tc\ and the distance to the nearest protostar for those
starless cores in the clusters NGC1333 and IC348.  However, there is a
significant negative correlation between \tc\ and the distance to the
nearest class 0 and/or I protostar for those starless cores in the
field (i.e. starless cores further from protostars are colder).  It is
not at all certain that this result contradicts the prediction of
\citet{Stamatellos07}, because we cannot say that nearby protostars
heat starless cores, just that warmer cores are found near protostars.

\subsection{\tc\ vs. Cloud Temperature} \label{TGTDTCO}

Here we compare the temperatures of starless cores with the
temperature of the material in which it is embedded.  We estimate the
line-of-sight averaged temperature of the dust (\td) \citep{Schnee08}
and the gas (\tco) \citep{Pineda08} in the Perseus molecular cloud and
compare it with the temperature derived from the \amm\ pointings.  The
resolution of the \td\ and \tco\ maps are both $\sim$40\arcsec, which
is roughly comparable to the 31\arcsec\ resolution of the ammonia
data.  However, both \td\ and \tco\ are imperfect measures of cloud
temperature.  Because $^{12}$CO is optically thick (and perhaps
depleted) it is not sensitive to the colder and denser cloud material
in which the starless cores are embedded.  The dust temperature
determination is emission-weighted, and at the FIR wavelengths (70 and
160 \micron) of the Spitzer maps the emission is dominated by warm
dust along the line of sight, and therefore is less sensitive to the
colder and denser clumps in which cores form.

Nevertheless, one might expect that the temperature of a starless core
could be affected, or predicted, by the temperature of the medium in
which is it embedded.  Starless cores are, by definition, externally
heated, so whatever is heating the cloud should also be heating the
embedded cores.  This analysis can be biased because the gas and dust
derived temperatures are warmer within the clusters than outside,
which is also true for \tc, so any derived correlations must take this
into account.  Figure \ref{TKINENVIRONMENT} and Tab.~\ref{CORRELTAB}
show that although both \tco\ and \td\ are significantly and
positively correlated with \tc\ when considering the entire population
of starless cores, this effect is entirely explained by the
cluster/field bias, and within each sub-population there is no
significant correlation.

\subsection{\tc\ vs.~Column Density} \label{TGNH}

The heating of starless cores is governed by the strength of the local
interstellar radiation field (ISRF).  Given that there are no O or B
stars within the Perseus molecular cloud, the ISRF seen by the cores
should depend largely on dust shielding but not on the distance to any
special source of external radiation.  Because we are unable to
measure the volume-averaged shielding around each starless core, we
use as a proxy the column density along the line of sight.  Based on
this simple analysis, one would predict that \tc\ would be
anti-correlated with column density.

On the other hand, for starless cores that are in pressure
equilibrium, the inward pressures due to the weight of the Perseus
molecular cloud and the self-gravity of the cores will be counter
balanced by the internal gas pressure from thermal and non-thermal
motion.  The external pressure on the starless cores from the
molecular cloud is proportional to the square of the column density
($P_{\rm cloud} \propto A_V^2$) \citep{Bertoldi92, Lada08} and the
internal gas pressure from thermal motions is proportial to \tc\
($P_{\rm thermal} \propto \sigma_{\rm T}^2 \propto$ \tc).  From a
balance of pressure analysis, one would expect a positive correlation
between \tc\ and column density, though this would not necessarily be
a strong effect because outward pressure is also supplied by
non-thermal motions and magnetic fields.  We calculate that for the
starless cores in our sample the inward pressure of the weight of the
molecular cloud is roughly a factor of 2 weaker than their
self-gravity.

Here we compare the temperatures of starless cores with the column
density of molecular cloud material along the line of sight to each
core.  We estimate the cloud column density using far-infrared dust
emission \citep{Schnee08} and near-infrared extinction
\citep{Ridge06}.  Although extinction may be the most secure estimator
of column density in a molecular cloud \citep{Goodman08}, the
40\arcsec\ resolution of the emission-derived column density map
compares favorably with the 5\arcmin\ resolution of the near-infrared
extinction map.  As with the cloud temperature, there is an inherent
bias in that the column density in the clusters is higher and the core
temperatures are higher, so we consider the field and cluster starless
cores separately.  Figure \ref{TKINENVIRONMENT} and Table
\ref{CORRELTAB} show that there is a significant positive correlation
between extinction-derived column density and \tc\ when considering
just the starless cores in the field and just the starless cores in
the clusters, as well as for the field+cluster ensemble.  There is no
significant trend between \tc\ and the dust emission-derived column
density for field or cluster cores.

One should keep in mind that the emission-derived column
density derived from dust emission loses accuracy at values of $A_V >$
a few, so it is not so surprising to get different results depending
on how the column density is estimated \citep{Schnee06}.  The dust
temperature derived from FIR maps is emission-weighted, and the column
density derivation depends on the dust temperature, so the column
density along lines of sight with significant variations in
temperature and density is both systemically underestimated and
subject to large scatter.  This effect could wash out any true
correlations between the column density of the cloud material and the
temperature of the core embedded in it.
  
In a numerical simulation of embedded starless cores,
\citet{Stamatellos03} show that the temperature profile of a core
shielded by $A_V$ = 5 material (a typical value for $A_V$ seen towards
the starless cores in Perseus) is not significantly different from a
naked core, while deeply embedded cores ($A_V$ = 20) are noticably
colder than naked cores.  In addition, we expect that the column
density along the line of sight is a poor indicator of volumetric
shielding due to the porosity of molecular clouds like Perseus.  The
visual impression given by two-dimensional maps and
position-position-velocity cubes of nearby molecular clouds is one of
non-relaxed structures with many filaments, voids and shells,
supporting this view.  The observational evidence supports the idea
that the shielding provided by higher \AV\ material in Perseus does
not lead to lower temperatures of the embedded starless cores.  The
opposite effect is observed, which may be explained by the greater
thermal support needed to balance the increased pressure on the cores
by the weight of the molecular cloud.

\subsection{\tc\ vs.~\sigamm\ and \sigco} 
\label{TGSIG}

Clumpiness formed by turbulence in a molecular cloud is known to
significantly affect the temperature of the cloud material, with
higher turbulence leading to greater porosity and penetration by the
ISRF, and therefore higher temperatures \citep{Bethell04}.  However,
\citet{Stamatellos07} find that the densest portions of the cloud
(where starless cores are found) are primarily heated by long
wavelength radiation that is not effectively shielded by the clumpy
structure.  It is therefore not obvious if one would expect starless
core temperatures to be correlated with turbulence.

Here we examine whether turbulence affects the temperature of the
starless cores in the Perseus molecular cloud.  We use the non-thermal
linewidth of \amm\ as a proxy for the turbulence within the cores, and
the total linewidth of $^{13}$CO as an estimate of the turbulence in
the surrounding medium (the measured $^{13}$CO linewidth is roughly a
factor of 10 higher than the thermal linewidth).  As before, we
consider field and cluster starless cores separately to avoid
confusing a correlation between \tc\ and turbulence with a correlation
between \tc\ and the field vs.~cluster environment.

\begin{figure}
\epsscale{1.0}
\plotone{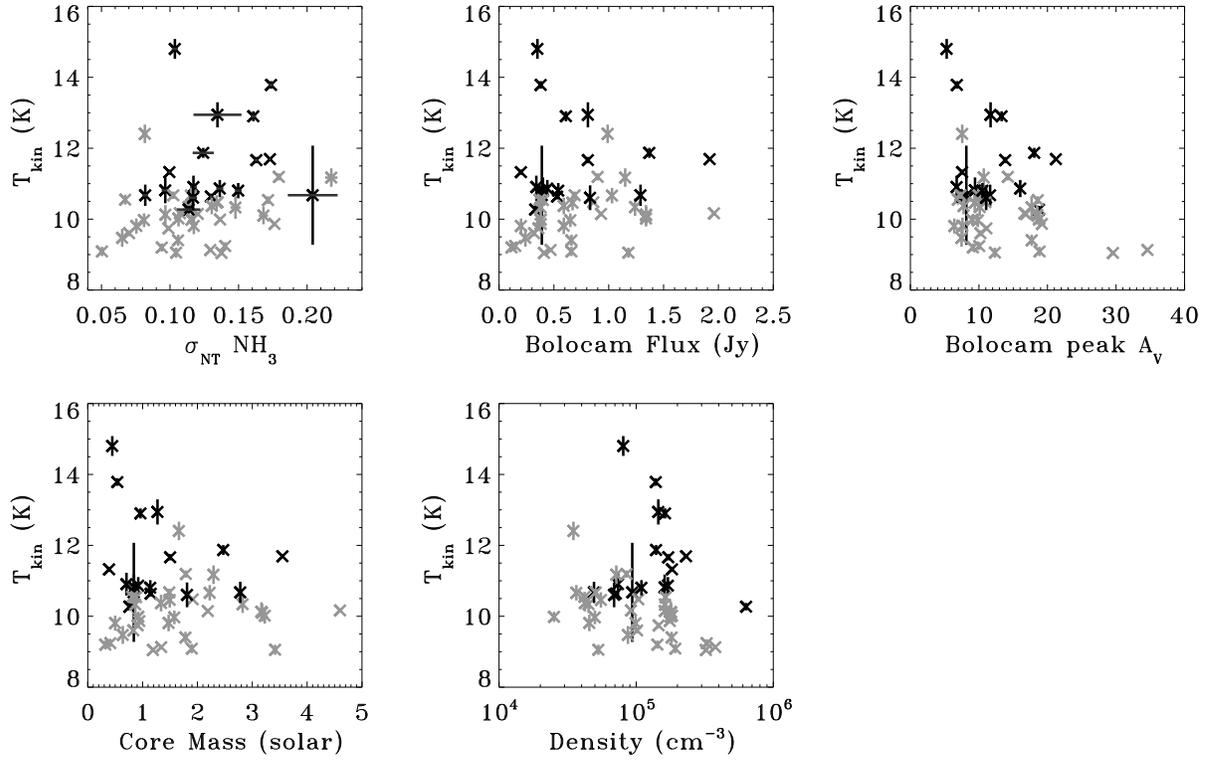}
\caption{\tc\ plotted against properties of the starless cores: (top
  left) non-thermal line width (1-D velocity dispersion) of \amm, (top
  middle) Bolocam Flux, (top right) peak \AV, (bottom left) core mass
  and (center) average volume density.  Starless cores in NGC1333 and
  IC348 are plotted in black, and field starless cores are plotted in
  grey.  The 1$\sigma$ errors in the fit of \tc\ and non-thermal
  linewidth are plotted for each core. \label{TKININTERNAL}}
\end{figure}

Figure \ref{TKININTERNAL} and Table \ref{CORRELTAB} show that there is
no correlation between the non-thermal linewidth of \amm\ and \tc\
when considering the field and cluster popoluations of starless cores
separately, but there is a significant positive correlation for the
ensemble.  This implies that those starless cores in Perseus are more
turbulent when found in clusters than in the field, but the turbulence
itself does not necessarily lead to higher temperatures.  Our
expectation of starless core geometry is that they are in some sense
``relaxed'' structures and therefore not porous, regardless of the
level of internal turbulence.  CO linewidth does not correlate with
\tc\ for those cores in clusters, but it is significantly positively
correlated with \tc\ for field starless cores, as shown in Figure
\ref{TKINENVIRONMENT}.  One interpretation of of the field/cluster
difference is that within clusters the CO linewidth is likely to be
affected strongly by the ongoing star formation, while the CO
linewidth in the field better traces the cloud geometry, so only in
the field does a larger CO linewidth imply better penetration by the
ISRF.

\subsection{\tc\ vs.~1.1mm Continuum Derived Properties}

It might be supposed that the temperatures of starless cores are
partly determined by the internal properties of the cores themselves,
and not just by their environment.  Here we compare \tc\ with 1.1mm
Bolocam flux, the peak column density (expressed in $A_V$), the mass
and the average volume density derived from the 1.1mm map and
\amm-derived temperature of each core.

The peak \AV\ is derived from the peak 1.1mm flux ($S_{\nu}^{beam}$)
of each core and the \amm-derived temperature.
\begin{equation}
N(\mathrm{H_2}) = \frac{S_{\nu}^{beam}}{\Omega_{beam} \mu_{H_2} m_H \kappa_{\nu} B_{\nu}(T_{\rm kin})}, \label{aveq}
\end{equation}
with $N($H$_2)/A_V = 0.94 \times 10^{21}$ cm$^{2}$~mag$^{-1}$
\citep{Frerking82}.  Here $\Omega_{beam}$ is the beam solid angle,
$m_H$ is the mass of hydrogen, $\kappa_{1.1mm} = 0.0114$ cm$^2$
g$^{-1}$ is the dust opacity per gram of gas interpolated from Table~1
column~5 of \citet{Ossenkopf94} for dust grains with thin ice mantles,
$B_{\nu}$ is the Planck function, and $\mu_{H_2}=2.8$ is the mean
molecular weight per H$_2$ molecule.  A gas to dust mass ratio of 100
is included in $\kappa_{1.1mm}$.

The core mass is derived from the total 1mm flux ($S_{\nu}$) and \tc\
using the equation,
\begin{equation}
M = \frac{d^2 S_{\nu}}{B_{\nu}(T_{\rm kin}) \kappa_{\nu}}, \label{masseq}
\end{equation}
where $\kappa_{\nu}$ is the dust opacity, $d$ is the distance to
Perseus (250 pc).  The density is derived from the core mass and the
deconvolved average FWHM size ($D$) of the core ($n = 6M/\pi D^3
\mu_p$), where $\mu_p = 2.33$ is the mean molecular weight per
particle, as explained in \citet{Enoch06}.

When deriving the core mass, peak column density and average volume
density from the 1.1mm continuum emission, we use the gas temperature
(\tc) in the calculation, under the assumption that the gas and dust
temperatures are coupled.  Although this assumption is valid for the
densities ($n > 2 \times 10^4$ cm$^{-3}$) derived for these cores
\citep{Goldsmith01, Galli02}, we are introducing a potential bias in
our analysis because \tc\ and the derived mass are not independent.
It would be preferable to use the T$_{\rm dust}$ rather than \tc\ to
determine the 1.1mm-derived core properties, but this requires mapping
the dust emission from each core at several wavelengths, sampling both
the peak of the spectral energy distribution (SED) as well as the
Rayleigh-Jeans tail to derive the dust temperature, column desnity and
emissivity spectral index profiles.  Existing surveys of Perseus
\citep{Enoch06, Kirk06} have only sampled the Rayleigh-Jeans portion
of the SED, so we do not have accurate estimates of the dust
temperatures within the starless cores, and we are forced to use the
gas temperature in our calculations.

Note that some of the Bolocam cores, whose numbers are given in Table
\ref{COREPROPTAB} (Per-Bolo 24, 75, 100 and 122), have been observed
in multiple GBT pointings with each position corresponding to
different sub-condensations within the core.  Each of these pointings
is assigned the same 1.1mm total and peak flux and dimensions as that
of the entire Bolocam core, and the peak $A_V$, mass and density are
all derived from the overall core Bolocam flux measurements and the
\tc\ derived for the sub-core.  An alternative analysis would have
been to break up each Bolocam core with more than one \amm\ pointing
into sub-cores and assign different fractions of the total 1.1mm flux
to each portion, but the procedure to do the division would be
somewhat arbitrary, just as the decision of which cores should be
observed only once and which in multiple positions was not based on an
absolute prescription.  Since neither analysis is obviously the
correct one, we decided to go with the simpler of the two and
double-count the 1.1mm properties of the four Bolocam cores with more
than one \amm\ observation.

In Fig.~\ref{TKININTERNAL} and Table \ref{CORRELTAB} we compare \tc\
with the 1.1mm flux and properties derived from the \citet{Enoch06}
Bolocam map of Perseus.  We find no correlation between \tc\ and the
1.1mm flux, core mass, average volume density or peak \AV\ for those
cores in IC348 and NGC1333.  On the other hand, for starless cores in
the field there are significant negative correlations between \tc\ and
the peak \AV\ and average volume density, and a positive correlation
between \tc\ and the Bolocam flux.  If we assume that starless cores
are effectively self-shielded from the ISRF because they are not
porous, then we would expect higher peak column densities and volume
densities to correspond to higher volumetric shielding and lower
internal temperatures.  It is not obvious why the self-shielding
argument does not hold for those starless cores in NGC1333 and IC348.
The connection between 1.1mm flux and \tc\ may be simply that for a
given mass a warmer core will produce more emission, but this argument
does not explain why the positive correlation is not seen in cluster
starless cores.

\section{Conclusions} \label{CONCLUSIONS}

In this paper we investigate the temperature distribution of starless
cores in Perseus.  Although, the range of temperatures derived from
\amm\ (1,1) and (2,2) pointing is quite narrow, from $\sim 9-15$ K,
there are several interesting correlations and lack of correlations.

The location of a starless core does affect its temperature.  We find
that those cores inside the two main star-forming regions in Perseus
(IC348 and NGC1333) are warmer than those not in clusters.  Although
there is no correlation between \tc\ and the distance to the nearest
class 0/I protostar for those starless cores in clusters, starless
cores in the field are colder the further away they are from
protostars.

Certain properties of the molecular cloud are related to the
temperatures of the starless cores embedded in it.  \tc\ is higher
when the extinction-derived column density of the molecular cloud is
larger, despite the additional shielding from the ISRF that the higher
column density implies, which may be due to the increased pressure on
the cores from the molecular cloud in higher column density regions.
This correlation is not seen when column density along the line of
sight in Perseus is measured from far-infrared dust emission.  After
controlling for the cluster/field temperature difference, we find that
\tc\ does not depend on the temperature of the molecular cloud, as
measured by CO or dust emission.  The clusters have higher values of
\tco\ and \td\ on average, so presumably the same processes that are
heating the clusters are also heating the cores within them.  Those
starless cores outside NGC1333 and IC348 are warmer in regions with
higher $^{13}$CO linewidth, suggesting the the turbulence causing the
high linewidths makes the cloud more porous to the ISRF, which in turn
heats the embedded cores.  There is no significant correlation between
$\sigma_{CO}$ and \tc\ for those cores in the clusters.

Some internal properties of starless cores are correlated with their
temperature. For those starless cores in the field, total 1.1mm flux
is higher in cores that are warmer, and peak column density and
average volume density are higher in cores that are colder.  A higher
temperature naturally leads to more 1.1mm flux for a given core mass,
and the correlations between \tc\ and peak core column density and
volume density can be explained by the amount of shielding from the
ISRF that the additional material provides.  However, it is surprising
that none of these correlations exist for starless cores inside the
clusters.  Support was provided to MLE by NASA through the Spitzer
Space Telescope Fellowship Program.

\acknowledgements

We thank an anonymous referee for comments that have improved this
paper.  SS acknowledges support from the Owens Valley Radio
Observatory, which is supported by the National Science Foundation
through grant AST 05-40399.  ER was supported by an NSF Astronomy and
Astrophysics Postdoctoral Fellowship (AST-0502605) and a Discovery
Grant from NSERC of Canada.  JBF is supported by a grant from the NRAO
Student Observing Support Program (GSSP06-0015)

\clearpage

\begin{deluxetable}{ccc|cccccc} 
\tablewidth{0pt}
\tabletypesize{\scriptsize}
\tablecaption{Internal Properties of Starless Cores \label{COREPROPTAB}}
\tablehead{
 \colhead{RA\tablenotemark{1}}         & 
 \colhead{Dec\tablenotemark{2}}        &
 \colhead{\tc\tablenotemark{3}}        &
 \colhead{\sigamm\tablenotemark{4}}    &
 \colhead{1.1mm Flux\tablenotemark{5}} &
 \colhead{Peak \AV\tablenotemark{6}}   & 
 \colhead{Mass\tablenotemark{7}}       &
 \colhead{Density\tablenotemark{8}}    &
 \colhead{Number\tablenotemark{9}}     \\ 
 \colhead{J2000}        & \colhead{J2000} &
 \colhead{K}            & \colhead{\kms}  & 
 \colhead{Jy}           & \colhead{mag}   &
 \colhead{M$_{solar}$}  & \colhead{\ccm}  &
 \colhead{}}
\startdata
3:25:07.8 & 30:24:22 &  9.2 & 0.09 & 0.1 &  9.1 & 0.3 & 1.4E+05 &   1 \\
3:25:09.7 & 30:23:53 &  9.2 & 0.14 & 0.1 & 10.1 & 0.4 & 3.3E+05 &   2 \\
3:25:10.1 & 30:44:41 & 10.0 & 0.12 & 0.4 &  9.0 & 0.9 & 2.5E+04 &   3 \\
3:25:17.1 & 30:18:53 & 10.3 & 0.11 & 1.2 & 10.1 & 2.8 & 4.5E+04 &   4 \\
3:25:26.9 & 30:21:53 &  9.1 & 0.10 & 1.2 & 12.3 & 3.4 & 5.3E+04 &   6 \\
3:25:46.1 & 30:44:11 & 11.2 & 0.18 & 0.9 & 14.2 & 1.8 & 8.5E+04 &  11 \\
3:25:48.8 & 30:42:24 &  9.1 & 0.13 & 0.5 & 34.6 & 1.3 & 3.8E+05 &  13 \\
3:25:50.6 & 30:42:02 &  9.0 & 0.14 & 0.4 & 29.6 & 1.2 & 3.2E+05 &  14 \\
3:27:02.1 & 30:15:08 & 10.4 & 0.13 & 0.6 &  7.8 & 1.3 & 4.2E+04 &  19 \\
3:27:28.9 & 30:15:04 & 10.7 & 0.10 & 0.7 &  7.3 & 1.5 & 5.1E+04 &  20 \\
3:27:40.0 & 30:12:13 & 10.5 & 0.17 & 0.4 & 18.6 & 0.9 & 1.6E+05 &  23 \\
3:27:55.9 & 30:06:18 &  9.1 & 0.05 & 0.7 & 18.9 & 1.9 & 1.9E+05 &  24 \\
3:28:05.5 & 30:06:19 &  9.4 & 0.11 & 0.7 & 17.7 & 1.8 & 1.8E+05 &  24 \\
3:28:32.4 & 31:04:43 & 10.6 & 0.13 & 0.5 &  7.6 & 1.1 & 7.1E+04 &  26 \\
3:28:33.4 & 30:19:35 & 10.7 & 0.12 & 1.0 & 10.0 & 2.2 & 3.7E+04 &  27 \\
3:28:39.5 & 31:18:35 & 11.7 & 0.17 & 1.9 & 21.2 & 3.6 & 2.3E+05 &  31 \\
3:28:48.5 & 31:16:03 & 11.3 & 0.10 & 0.2 &  7.6 & 0.4 & 1.8E+05 &  35 \\
3:28:52.2 & 31:18:08 & 13.8 & 0.17 & 0.4 &  6.8 & 0.5 & 1.4E+05 &  37 \\
3:29:18.5 & 31:25:13 & 11.9 & 0.12 & 1.4 & 18.1 & 2.5 & 1.4E+05 &  53 \\
3:29:22.5 & 31:36:24 &  9.8 & 0.12 & 0.6 &  7.6 & 1.5 & 4.6E+04 &  56 \\
3:29:25.8 & 31:28:17 & 10.3 & 0.11 & 0.3 & 18.8 & 0.8 & 6.3E+05 &  58 \\
3:30:24.1 & 30:27:39 & 10.5 & 0.07 & 0.4 &  7.0 & 0.8 & 4.2E+04 &  61 \\
3:30:45.6 & 30:52:36 & 10.5 & 0.13 & 0.7 &  9.4 & 1.5 & 5.5E+04 &  63 \\
3:30:50.5 & 30:49:17 &  9.8 & 0.08 & 0.2 &  6.4 & 0.5 & 9.9E+04 &  64 \\
3:32:26.9 & 30:59:11 & 10.5 & 0.13 & 0.9 & 11.0 & 1.9 & 1.0E+05 &  67 \\
3:32:44.1 & 31:00:01 & 10.2 & 0.11 & 2.0 & 16.7 & 4.6 & 9.2E+04 &  70 \\
3:32:51.3 & 31:01:48 & 11.2 & 0.22 & 1.1 & 10.7 & 2.3 & 7.2E+04 &  71 \\
3:32:57.0 & 31:03:21 & 10.1 & 0.12 & 0.9 & 16.8 & 2.2 & 1.6E+05 &  72 \\
3:33:02.0 & 31:04:33 & 10.0 & 0.14 & 0.4 & 18.8 & 0.9 & 1.8E+05 &  74 \\
3:33:04.3 & 31:04:57 &  9.9 & 0.18 & 0.4 & 19.2 & 0.9 & 1.8E+05 &  75 \\
3:33:06.3 & 31:06:26 & 10.3 & 0.15 & 0.4 & 17.7 & 0.9 & 1.6E+05 &  75 \\
3:33:25.2 & 31:05:35 &  9.7 & 0.10 & 0.4 & 11.1 & 0.9 & 1.5E+05 &  82 \\
3:40:49.5 & 31:48:35 & 12.4 & 0.08 & 1.0 &  7.6 & 1.7 & 3.5E+04 &  89 \\
3:41:40.2 & 31:58:05 &  9.5 & 0.07 & 0.2 &  7.5 & 0.6 & 8.7E+04 &  92 \\
3:41:46.0 & 31:57:22 &  9.6 & 0.07 & 0.3 & 10.2 & 0.8 & 1.0E+05 &  94 \\
3:43:44.0 & 32:02:52 & 12.9 & 0.13 & 0.8 & 11.7 & 1.3 & 1.5E+05 & 100 \\
3:43:45.5 & 32:01:44 & 14.0 & 0.10 & 0.3 &  5.3 & 0.4 & 8.0E+04 & 101 \\
3:43:45.8 & 32:03:11 & 11.7 & 0.16 & 0.8 & 13.8 & 1.5 & 1.7E+05 & 100 \\
3:43:57.8 & 32:04:06 & 12.9 & 0.16 & 0.6 & 13.3 & 1.0 & 1.6E+05 & 105 \\
3:44:05.1 & 32:00:28 & 10.7 & 0.20 & 0.4 &  8.2 & 0.8 & 9.4E+04 & 109 \\
3:44:05.3 & 32:02:05 & 10.9 & 0.14 & 0.4 & 16.0 & 0.9 & 1.7E+05 & 110 \\
3:44:14.6 & 31:57:59 & 10.6 & 0.12 & 0.8 & 11.1 & 1.8 & 6.9E+04 & 111 \\
3:44:36.4 & 31:58:40 & 10.8 & 0.15 & 0.5 & 10.5 & 1.1 & 1.1E+05 & 115 \\
3:44:48.8 & 32:00:29 & 10.8 & 0.10 & 0.4 &  9.4 & 0.8 & 1.6E+05 & 117 \\
3:44:56.1 & 32:00:32 & 10.9 & 0.12 & 0.3 &  6.8 & 0.7 & 7.4E+04 & 118 \\
3:45:15.9 & 32:04:49 & 10.7 & 0.08 & 1.3 & 11.6 & 2.8 & 4.9E+04 & 119 \\
3:47:33.5 & 32:50:55 & 10.0 & 0.08 & 0.7 &  9.9 & 1.6 & 5.0E+04 & 121 \\
3:47:38.6 & 32:52:19 & 10.1 & 0.17 & 1.3 & 18.5 & 3.2 & 1.8E+05 & 122 \\
3:47:39.7 & 32:53:57 & 10.1 & 0.10 & 1.3 & 18.5 & 3.2 & 1.8E+05 & 122 \\
3:47:39.8 & 32:53:34 & 10.0 & 0.11 & 1.3 & 18.8 & 3.2 & 1.8E+05 & 122
\enddata
\tablenotetext{1}{J2000 position of \amm\ pointing, $^2$J2000 position
of \amm\ pointing, $^3$Kinetic temperature of core derived from \amm\
(1,1) and (2,2) spectra, $^4$Non-thermal 1-D velocity dispersion of
\amm\ spectra, $^5$Total 1.1mm flux of the Bolocam core, $^6$Peak
column density of the core derived from the peak 1.1mm flux and \tc,
$^7$Total mass of Bolocam core derived from the integrated 1.1mm flux
and \tc, $^8$Average density of free particles of Bolocam core derived
from its mass and size, $^9$Per-Bolo number taken from Table 4 in
\citet{Enoch08}}
\end{deluxetable}

\begin{deluxetable}{ccc|cccccccc} 
\tablewidth{0pt}
\tabletypesize{\scriptsize}
\tablecaption{Environment of Starless Cores \label{COREENVIRTAB}}
\tablehead{
 \colhead{RA\tablenotemark{1}}	               &
 \colhead{Dec\tablenotemark{2}}	               &
 \colhead{\tc\tablenotemark{3}}                &
 \colhead{\td\tablenotemark{4}}                & 
 \colhead{\tco\tablenotemark{5}}               &
 \colhead{NIR \AV\tablenotemark{6}}            &
 \colhead{FIR \AV\tablenotemark{7}}            &
 \colhead{\sigco\tablenotemark{8}}             &    
 \colhead{Distance Class 0/I\tablenotemark{9}} &
 \colhead{In a Cluster?\tablenotemark{10}}     \\
 \colhead{J2000}              & \colhead{J2000}         &
 \colhead{K}                  & \colhead{K}             & 
 \colhead{K}                  & \colhead{mag}           &
 \colhead{mag}                & \colhead{\kms}          &
 \colhead{pc}                 & \colhead{}}
\startdata
3:25:07.8 & 30:24:22 &  9.2 & 14.6 & 11.6 &  5.0 &  2.6 & 0.5 & 1.51 & N \\
3:25:09.7 & 30:23:53 &  9.2 & 14.6 & 11.2 &  5.0 &  2.6 & 0.6 & 1.50 & N \\
3:25:10.1 & 30:44:41 & 10.0 & 15.2 & 12.9 &  5.1 &  4.0 & 0.6 & 0.19 & N \\
3:25:17.1 & 30:18:53 & 10.3 & 12.9 & 12.0 &  5.0 &  7.2 & 0.7 & 1.28 & N \\
3:25:26.9 & 30:21:53 &  9.1 & 12.9 & 12.7 &  3.7 &  6.7 & 0.5 & 1.20 & N \\
3:25:46.1 & 30:44:11 & 11.2 & 15.5 & 11.0 &  5.5 &  5.1 & 0.7 & 0.12 & N \\
3:25:48.8 & 30:42:24 &  9.1 & 14.3 & 11.2 &  4.8 &  5.4 & 0.7 & 0.20 & N \\
3:25:50.6 & 30:42:02 &  9.0 & 12.5 & 10.6 &  4.8 & 10.4 & 0.7 & 0.24 & N \\
3:27:02.1 & 30:15:08 & 10.4 & 13.5 & 13.1 &  4.8 &  3.8 & 0.6 & 0.39 & N \\
3:27:28.9 & 30:15:04 & 10.7 & 14.9 &  NaN &  9.5 &  2.9 & NaN & 0.16 & N \\
3:27:40.0 & 30:12:13 & 10.5 & 17.4 &  NaN &  8.8 &  2.8 & NaN & 0.04 & N \\
3:27:55.9 & 30:06:18 &  9.1 & 13.7 &  NaN &  3.1 &  3.1 & NaN & 0.17 & N \\
3:28:05.5 & 30:06:19 &  9.4 & 13.7 &  NaN &  2.9 &  3.1 & NaN & 0.16 & N \\
3:28:32.4 & 31:04:43 & 10.6 & 12.8 & 18.8 &  5.4 &  8.3 & 1.1 & 0.14 & Y \\
3:28:33.4 & 30:19:35 & 10.7 & 13.0 &  NaN &  4.5 &  6.0 & NaN & 0.90 & N \\
3:28:39.5 & 31:18:35 & 11.7 & 12.3 & 18.6 &  7.8 & 19.1 & 0.8 & 0.05 & Y \\
3:28:48.5 & 31:16:03 & 11.3 & 16.0 & 16.2 &  8.5 &  0.0 & 0.9 & 0.15 & Y \\
3:28:52.2 & 31:18:08 & 13.8 & 15.0 & 19.4 &  9.2 & 10.7 & 1.3 & 0.10 & Y \\
3:29:18.5 & 31:25:13 & 11.9 &  NaN & 49.0 &  6.5 &  NaN & 0.7 & 0.13 & Y \\
3:29:22.5 & 31:36:24 &  9.8 & 15.2 & 15.5 &  5.9 &  2.9 & 0.5 & 0.23 & N \\
3:29:25.8 & 31:28:17 & 10.3 & 15.6 & 23.3 &  5.5 &  0.0 & 1.2 & 0.14 & Y \\
3:30:24.1 & 30:27:39 & 10.5 & 14.2 &  NaN &  7.0 &  5.5 & NaN & 0.16 & N \\
3:30:45.6 & 30:52:36 & 10.5 & 13.3 & 12.4 &  2.7 &  4.8 & 0.9 & 0.74 & N \\
3:30:50.5 & 30:49:17 &  9.8 & 13.5 & 11.6 &  2.1 &  2.8 & 0.8 & 0.54 & N \\
3:32:26.9 & 30:59:11 & 10.5 & 13.4 & 13.5 & 10.3 &  9.0 & 0.7 & 0.23 & N \\
3:32:44.1 & 31:00:01 & 10.2 & 12.2 & 12.8 & 10.0 & 21.7 & 0.9 & 0.30 & N \\
3:32:51.3 & 31:01:48 & 11.2 & 12.3 & 13.7 &  8.8 & 20.3 & 0.9 & 0.36 & N \\
3:32:57.0 & 31:03:21 & 10.1 & 12.4 & 14.4 &  8.5 & 15.4 & 0.8 & 0.39 & N \\
3:33:02.0 & 31:04:33 & 10.0 & 12.5 & 13.9 &  8.5 & 16.7 & 0.9 & 0.27 & N \\
3:33:04.3 & 31:04:57 &  9.9 & 12.5 & 14.0 &  8.5 & 15.3 & 0.8 & 0.23 & N \\
3:33:06.3 & 31:06:26 & 10.3 & 12.7 & 13.3 &  8.5 & 13.3 & 0.9 & 0.14 & N \\
3:33:25.2 & 31:05:35 &  9.7 & 12.5 & 15.8 &  8.6 & 19.1 & 0.9 & 0.10 & N \\
3:40:49.5 & 31:48:35 & 12.4 & 16.1 & 15.4 &  7.7 &  9.1 & 0.9 & 0.42 & N \\
3:41:40.2 & 31:58:05 &  9.5 & 16.3 & 21.3 &  8.4 &  7.5 & 0.6 & 1.09 & N \\
3:41:46.0 & 31:57:22 &  9.6 & 15.6 & 19.7 &  8.4 & 10.3 & 0.7 & 1.09 & N \\
3:43:44.0 & 32:02:52 & 12.9 &  NaN & 21.8 &  9.5 &  NaN & 0.6 & 0.11 & Y \\
3:43:45.5 & 32:01:44 & 14.0 & 14.6 & 20.7 & 10.1 & 15.8 & 0.5 & 0.14 & Y \\
3:43:45.8 & 32:03:11 & 11.7 & 15.1 & 21.4 &  8.7 & 21.3 & 0.6 & 0.07 & Y \\
3:43:57.8 & 32:04:06 & 12.9 & 15.4 & 27.0 &  7.4 & 19.0 & 0.8 & 0.08 & Y \\
3:44:05.1 & 32:00:28 & 10.7 & 14.7 & 21.9 &  9.0 & 14.8 & 0.8 & 0.13 & Y \\
3:44:05.3 & 32:02:05 & 10.9 &  NaN & 23.5 &  9.0 &  NaN & 0.6 & 0.06 & Y \\
3:44:14.6 & 31:57:59 & 10.6 &  NaN & 17.0 &  7.9 &  NaN & 0.7 & 0.16 & Y \\
3:44:36.4 & 31:58:40 & 10.8 & 15.3 & 16.8 &  7.2 &  9.1 & 0.8 & 0.22 & Y \\
3:44:48.8 & 32:00:29 & 10.8 & 14.8 & 16.9 &  5.9 & 12.5 & 0.6 & 0.10 & Y \\
3:44:56.1 & 32:00:32 & 10.9 & 14.9 & 18.6 &  5.4 & 10.0 & 0.5 & 0.20 & Y \\
3:45:15.9 & 32:04:49 & 10.7 & 14.3 & 24.5 &  5.2 & 12.3 & 0.5 & 0.55 & Y \\
3:47:33.5 & 32:50:55 & 10.0 & 14.0 & 16.5 &  5.8 &  6.5 & 0.5 & 0.14 & N \\
3:47:38.6 & 32:52:19 & 10.1 & 13.4 & 16.0 &  7.6 & 10.7 & 0.7 & 0.06 & N \\
3:47:39.7 & 32:53:57 & 10.1 & 12.7 & 14.2 &  7.6 & 11.9 & 0.6 & 0.16 & N \\
3:47:39.8 & 32:53:34 & 10.0 & 12.4 & 14.8 &  7.6 & 14.8 & 0.6 & 0.13 & N \\
\enddata
\tablenotetext{1}{J2000 position of \amm\ pointing, $^2$J2000 position
of \amm\ pointing, $^3$Kinetic temperature of core derived from \amm\
(1,1) and (2,2) spectra, $^4$Dust temperature of the Perseus molecular
cloud derived from FIR emission, $^5$Gas temperature of the Perseus
molecular cloud derived from $^{12}$CO (1--0) emission, $^6$Column
density of the Perseus molecular cloud derived from NIR extinction,
$^7$Column density of the Perseus molecular cloud derived from FIR
emission, $^8$ 1-D velocity dispersion of $^{13}$CO (1--0),
$^9$Distance to nearest Class 0/I protostar, $^{10}$Is the core within
a cluster (NGC1333 or IC348)?}
\end{deluxetable}

\begin{deluxetable}{lccccc} 
\tablewidth{0pt}
\tabletypesize{\scriptsize}
\tablecaption{\tc\ Correlations\label{CORRELTAB}}
\tablehead{
 \colhead{Property}                      & 
 \colhead{Location}                      &
 \colhead{r value\tablenotemark{1}}      &
 \colhead{t value\tablenotemark{2}}      &
 \colhead{confidence\tablenotemark{3}}   & 
 \colhead{DOF\tablenotemark{4}}}
\startdata
\td                       & field+cluster &  0.27 & 1.83 & 96.3    & 44 \\
\tco                      & field+cluster &  0.45 & 3.26 & $>$99.9 & 42 \\
NIR \AV                   & field+cluster &  0.47 & 3.66 & $>$99.9 & 48 \\
FIR \AV                   & field+cluster &  0.34 & 2.43 & 99.0    & 44 \\
\sigco                    & field+cluster &  0.20 & 1.31 & 90.1    & 42 \\
\sigamm                   & field+cluster &  0.28 & 2.06 & 97.8    & 48 \\
1.1mm Flux                & field+cluster &  0.11 & 0.73 & 76.6    & 48 \\
Peak \AV                  & field+cluster & -0.33 & 2.39 & 99.0    & 48 \\
Mass                      & field+cluster & -0.11 & 0.78 & 78.0    & 48 \\
Density                   & field+cluster & -0.18 & 1.25 & 89.1    & 48 \\
dist 0\tablenotemark{5}   & field+cluster & -0.45 & 2.29 & 98.3    & 21 \\
dist I\tablenotemark{5}   & field+cluster & -0.21 & 1.14 & 86.8    & 29 \\
dist 0/I\tablenotemark{5} & field+cluster & -0.29 & 1.71 & 95.1    & 33 \\[3pt]
\hline \\
\td                       & field only    &  0.16 & 0.93 & 82.0    & 31 \\
\tco                      & field only    &  0.04 & 0.20 & 57.8    & 25 \\
NIR \AV                   & field only    &  0.36 & 2.17 & 98.1    & 31 \\
FIR \AV                   & field only    &  0.16 & 0.91 & 81.5    & 31 \\
\sigco                    & field only    &  0.50 & 2.90 & 99.6    & 25 \\
\sigamm                   & field only    &  0.27 & 1.55 & 93.5    & 31 \\
1.1mm Flux                & field only    &  0.35 & 2.06 & 97.6    & 31 \\
Peak \AV                  & field only    & -0.36 & 2.17 & 98.1    & 31 \\
Mass                      & field only    &  0.15 & 0.86 & 80.1    & 31 \\
Density                   & field only    & -0.54 & 3.58 & $>$99.1 & 31 \\
dist 0\tablenotemark{5}   & field only    & -0.61 & 2.32 & 97.7    &  9 \\
dist I\tablenotemark{5}   & field only    & -0.54 & 2.43 & 98.6    & 14 \\
dist 0/I\tablenotemark{5} & field only    & -0.51 & 2.42 & 98.7    & 17 \\[3pt]
\hline \\
\td                       & cluster only  &  0.06 & 0.19 & 57.4    & 11 \\
\tco                      & cluster only  &  0.12 & 0.46 & 67.4    & 15 \\
NIR \AV                   & cluster only  &  0.63 & 3.13 & 99.7    & 15 \\
FIR \AV                   & cluster only  &  0.39 & 1.41 & 90.7    & 11 \\
\sigco                    & cluster only  & -0.09 & 0.36 & 63.8    & 15 \\
\sigamm                   & cluster only  &  0.14 & 0.56 & 70.8    & 15 \\
1.1mm Flux                & cluster only  & -0.03 & 0.12 & 54.7    & 15 \\
Peak \AV                  & cluster only  & -0.25 & 0.99 & 83.1    & 15 \\
Mass                      & cluster only  & -0.19 & 0.76 & 77.0    & 15 \\
Density                   & cluster only  & -0.19 & 0.75 & 76.8    & 15 \\
dist 0\tablenotemark{5}   & cluster only  & -0.26 & 0.86 & 79.5    & 10 \\
dist I\tablenotemark{5}   & cluster only  & -0.15 & 0.55 & 70.4    & 13 \\
dist 0/I\tablenotemark{5} & cluster only  & -0.18 & 0.69 & 74.9    & 14 \\
\enddata
\tablenotetext{1}{coefficient of correlation ($r$), $^2$t statistic:
$r/{\rm sqrt}[(1-r^2)/{\rm DOF}]$, $^3$\% confidence of correlation,
$^4$degrees of freedom, $^5$distance to a Class 0/I source, as
identified in \citet{Enoch08b} and plotted in Figure
\ref{TKINENVIRONMENT}}
\end{deluxetable}

\clearpage

\end{document}